\documentclass[%
 reprint,
 amsmath,amssymb,
 aps,
]{revtex4-1}

\usepackage{graphicx}% Include figure files
\usepackage{dcolumn}% Align table columns on decimal point
\usepackage{bm}% bold math
\begin{document}

\preprint{APS/123-QED}

\title{Free energy potential and temperature with information exchange}% Force line breaks with \\

%\collaboration{}%\noaffiliation

\author{Alessio Gagliardi, Aldo Di Carlo}
 \homepage{http://www.Second.institution.edu/~Charlie.Author}
\affiliation{ Department of Electronic Engineering, University of
Rome "Tor Vergata", CHOSE\\ Via del Politecnico 1, 00133, Rome,
Italy \\ gagliardi@ing.uniroma2.it,
Tel: (+39) 06 7259-7367, Fax: (+39) 06 7259-7939. \\
}%

\date{\today}% It is always \today, today,

\begin{abstract}
  In this paper we develop a generalized formalism for
  equilibrium thermodynamic systems when an information is shared
  between the system and the reservoir. The information results in
  a correction to the entropy of the system. This extension of the
  formalism requires a consistent generalization of the concept of
  thermodynamic temperature. We show that this extended equilibrium formalism
  includes also non-equilibrium conditions in steady state. By non-equilibrium conditions we mean here
  a non Boltzmann probability distribution within the phase space of the system.
  It is in fact possible to map non-equilibrium steady state in an equivalent
  system in equilibrium conditions (Boltzmann distribution) with generalized temperature and
  the inclusion of the information potential corrections. A simple model consisting in a
  single free particle is discussed as elementary application of the theory.
\begin{description}
\item[PACS numbers] May be entered using the \verb+\pacs{#1}+
command.
\end{description}
\end{abstract}

\maketitle

%\tableofcontents

In the last fifty years, a big effort has been devoted in trying
to extend the thermodynamic formalism beyond equilibrium
conditions~\cite{crooksthesis,jarzynski4,crooks4,boksen,muschik2,crooks2,crooks3,zubarev2,luzzi1}.
Similar studies have been carried out about the definition of a
non-equilibrium temperature~\cite{diventra,tempneq}. Moreover, our
civilization is going in the direction of diffuse information
sharing where information technology is ubiquitous. The inclusion
of information sharing effects within conventional thermodynamic
systems is then of paramount relevance \cite{sagawa2,sagawa3}.

In this work we present a formalism to extend the concept of free
energy when, beside heat or particles, the system exchanges also
information with the reservoir~\cite{alessio}. The extension of
the formalism requires a consistent generalization of the concept
of thermodynamic temperature. The sharing of information between
system and reservoir can be interpreted as the presence of a
Maxwell Demon which is able to reduce the internal entropy of the
system. The formalism can be also extended to include steady state non equilibrium systems
where the probability distribution (PD) within phase space does not follow Boltzmann distribution.

Information is the central quantity in our work and it is already
present within the definition of conventional entropy. Entropy $S$ is
an extensive quantity which measures the internal disorder of the
system: $S = -k \int p(\vec{x}) \ln \left ( p(\vec{x}) \right ) d\vec{x}$,
where $k$ is the Boltzmann constant and
$p(\vec{x})$ the probability of the system to be in the
$\vec{x}^{th}$ microstate. Every microstate configuration has a
well defined energy $\mathcal{E}(\vec{x})$, where
$\mathcal{E}(\vec{x})$ is the Hamiltonian of the system. The nature of $\vec{x}$ degrees of freedom
depend on the problem at hand, for example, for a gas, they represent the collection of particle
positions and momenta.

From the knowledge of the entropy function it is possible to
construct free energy potentials, which tell the amount of energy
within the system that can be effectively used to make work.
Helmoltz free energy is defined as: $F = \bar{E} - T_{\theta} S$,
with $\bar{E}$ average internal energy and $T_{\theta}$ the
thermodynamic temperature. The relation between
entropy and thermodynamic temperature is equal to: $\partial S / \partial \bar{E} = T^{-1}_{\theta}$.
We can define another temperature, kinetic temperature ($T_k$), related to
the average energy per degree of freedom $\bar{\xi}$ = $\bar{E}/N$, with $N$ number of degrees of freedom.
$\bar{\xi}$ $\propto$ $k T_{k}$.

The two temperatures, kinetic and thermodynamic, must be equal to
the reservoir temperature $T_0$ in equilibrium conditions. This is
assured by the equipartition theorem~\cite{huang} for the kinetic
temperature and by the equilibrium PD
for the thermodynamic temperature. In non equilibrium conditions,
when the PD is different from the Boltzmann distribution, the two
temperatures can be different and not equal to the reservoir
temperature. Even more important, Narayanan and
Sanistrava~\cite{sanistrava} have demonstrated for a Langevin
model that the differences between these three temperatures have
very deep meaning:
\begin{eqnarray}
     \Delta Q & \propto & T_{0} - T_{\theta} \quad and \quad  \Delta W  \propto  T_{\theta} - T_{k},
\label{tempeff}
\end{eqnarray}
where $\Delta Q$ and $\Delta W$ are the heat and mechanical work of
the system exchanged with the reservoir, respectively. A positive value
means that the system is absorbing energy, a negative that it is
giving energy to the reservoir. In equilibrium conditions the
fluxes are both zero.

The extension of conventional thermodynamic equilibrium formalism
to information based and non-equilibrium conditions make use of
informational theory concepts. This is possible thanks to the
deep link that exists between Shannon entropy, $H(\vec{x})$, and
thermodynamic entropy~\cite{cover}. The two are
formally equivalent, apart for the immaterial Boltzmann constant,
i.e. $H(\vec{x})$ = $S/k$. In Shannon entropy, $\vec{x}$
represents the set of possible vectors of symbols (words) that can
be emitted by an information source.

The use of information theory concepts in thermodynamics (and
viceversa) is not
new~\cite{jaynes,jaynes2,brill,merhav,bagci,landauer,jarzynski3}.
We extend previous works to include the effect of information
within both temperature and free energy, we will show that those
are the correct ingredients to include also the non-equilibrium
case in steady state.

In order to make a connection between information theory and
thermodynamics we start making a link between phase space in
statistical physics and the space of possible words of an
information source. The phase space is the set of configurations
(microstates) of a system considering all its relevant degrees of
freedom. The occupancy of different microstates is controlled by
the PD, $p(\vec{x})$. The PD defines which configurations of the
system are most likely or are forbidden. If a system is in contact
with an external reservoir with temperature $T_0$, then the
equilibrium PD, for a classical system, follows Boltzmann
distribution. The relation between phase space and entropy is
described by the fundamental postulate,
\begin{equation}\label{vol1}
    |\Omega(\vec{x})| = e^{\frac{S}{k}},
\end{equation}
where $|\Omega(\vec{x})|$ is the volume of the allowed microstates
within the phase space. The link between information theory and
thermodynamics is given by the Asymptotic equipartition principle
(AEP)~\cite{cover}. This states that if we have a source which is
emitting vector of symbols, then, in the limit of a vector of
infinite size (equivalent to a thermodynamic limit) only a
fraction collects almost all the probability. The so called
typical set has a volume which depends on the Shannon entropy of
the source, exactly in the same form as in
eq.~(\ref{vol1})~\cite{merhav}, $|\Omega(\vec{x})| = e^{H(\vec{x})}$.

This condition can be further generalized by introducing
correlation between vectors of random variables. Lets assume that
we have two vectors $\vec{x}$ (of a system $X$) and $\vec{y}$
(system $Y$) which are correlated. An outcome of
$\vec{y}$ gives information about the estimation of the state $\vec{x}$. This
means that the entropy of $X$ is reduced by this correlation. This reduction is quantified by the mutual
information, $I(\vec{x} \wedge \vec{y})$~\cite{kullback}:
\begin{equation}\label{inf1}
   I(\vec{x} \wedge \vec{y}) = \int \int p(\vec{x},\vec{y}) \ln
   \left (\frac{p(\vec{x},\vec{y})}{p(\vec{x})p(\vec{y})} \right )
   d\vec{x} d\vec{y}.
\end{equation}
The difference between the entropy of $X$ and mutual information
relates to the conditional entropy, $H(\vec{x}|\vec{y})$ =
$H(\vec{x})$ - $I( \vec{x} \wedge \vec{y})$, which is always
smaller than the full entropy $H(\vec{x})$.

The connection with thermodynamic systems depends on the
interpretation we give to $\vec{x}$ and $\vec{y}$. There are
fundamentally two cases. In both cases $\vec{x}$ collects all the
degrees of freedom of the system and $\vec{y}$ represents external
variables which are correlated with the system. In the first case
$\vec{y}$ represents a set of probes of the system $X$ that
measures its internal state. An outcome of
$\vec{y}$ allows to better estimates the microstate of the system
reducing its entropy. In this first case the reservoir
gains only information, but does
not perturb the Hamiltonian, i.e. $\mathcal{E}(\vec{x})$ is
independent of $\vec{y}$. This means that the average energy of
the system is not perturbed: $\bar{E}$ = $\int
p(\vec{x},\vec{y})\mathcal{E}(\vec{x}) d\vec{x}d\vec{y}$ = $\int
p(\vec{x})\mathcal{E}(\vec{x}) d\vec{x}$. On the
contrary the thermodynamic temperature and the entropy are
changed. This first case describes a system in equilibrium,
sharing information with the reservoir. This case will be called
"Information and equilibrium" ($Ieq$).

A second case occurs if $\vec{y}$ affects also the Hamiltonian, $\mathcal{E}(\vec{x},\vec{y})$. Then we expect that the average
energy $\bar{E}_{Neq}$ is different from the equilibrium value.
This second case, including a modification of the Hamiltonian,
represents a steady state non equilibrium condition ($Neq$).

To make a simple example of the two cases we assume a gas of heavy atoms measured sending photons.
The photons are then collected by a photo detector. The outcome of the sensor depends on the details of the photons collected ($\vec{y}$) which
are affected by many random sources, such as thermal noise, but are also correlated to the positions and momenta of the atoms ($\vec{x}$). If the photons
only negligible perturb the dynamics of the particles then the reservoir is only probing the system, but not affecting its Hamiltonian. On the contrary, with high energy
photons, we expect that the impacts between atoms and photons also perturb the dynamic of the system and then the Hamiltonian of the particles. The perturbation
is now genuinely driving it out of equilibrium.

When the system $X$ is controlled by two sets of random variables, $\vec{x}$ and $\vec{y}$, and by their joint probability,
$p(\vec{x}, \vec{y})$, the typical set must be replaced by the
joint typical set. The joint typical set is the set of typical
vectors $\vec{z} = (\vec{x}, \vec{y})$. The typical $\vec{z}$
vectors are formed by typical vectors $\vec{x}$ and $\vec{y}$
which are also jointly typical. In fact not all the typical
vectors within the $X$ and $Y$ phase spaces alone are also jointly
typical. For a rigorous mathematical definition of joint typical
set we refer to~\cite{cover}. The joint typical set is linked to
the joint entropy by a similar relation as for AEP~\cite{cover}, $|\Omega(\vec{x},\vec{y})| = e^{H(\vec{x},\vec{y})}$,
where the joint entropy is given by $H(\vec{x},\vec{y}) =
H(\vec{x}) + H(\vec{y}) - I(\vec{x} \wedge \vec{y})$.

However, $\vec{y}$ does not belong to the system $X$. The joint typical volume must be
normalized with respect to the volume of the typical set of $Y$:
\begin{eqnarray}
  \frac{|\Omega(\vec{x},\vec{y})|}{|\Omega(\vec{y})|} &=&
  |\Omega(\vec{x}|\vec{y})| =  \frac{e^{H(\vec{x},\vec{y})}}{e^{H(\vec{y})}} = e^{H(\vec{x})}e^{-I(\vec{x} \wedge \vec{y})}.
\label{vol5}
\end{eqnarray}
The entropy is substituted by conditional entropy $H(\vec{x} |
\vec{y})$ in both cases. $H(\vec{x})$ is the total entropy. The
external correlation with $\vec{y}$, whatever it
represents just a collection of information ($Ieq$) or a genuine perturbation ($Neq$), has the effect, considering the
fundamental postulate (eq.~\ref{vol1}), of compressing the volume
of the relevant phase space of $X$ by a factor $exp(-I(\vec{x} \wedge \vec{y}))$.

Generalized free energy potentials cannot be developed
irrespective of an equal generalization of thermodynamic
temperature related to conditional entropy:
\begin{equation}\label{thermo1}
    \frac{1}{kT_{\theta}} = \frac{\partial H(\vec{x}|\vec{y}))}{\partial \bar{E}}.
\end{equation}

In this paper we concentrate on a special case of eq. (\ref{thermo1}) which links entropy to a central quantity in estimation theory.
This particular generalization has been developed in the work of Narayanan and
Srinivasa~\cite{sanistrava}. We have further
extended their concept assuming also conditional probability which
includes the situation with information sharing and full non equilibrium.

This link between thermodynamic temperature and information theory
is given by the De Bruijn's identity. De Bruijn's identity states
that if we add an infinitesimal perturbation to the random
variable vector $\vec{x}$, i.e. $\vec{x}$ $\rightarrow$ $\vec{x}$
+ $\sqrt{\epsilon}$$\vec{z}$~\cite{footnote}, with $\epsilon$
$\rightarrow$ 0$^{+}$, then we get~\cite{cover}:
\begin{equation}\label{temp1}
    \frac{\partial H(\vec{x} + \sqrt{\epsilon}\vec{z})}{\partial
    \epsilon} = \frac{Tr[\mathbf{J}(\vec{x})]}{2}.
\end{equation}
In eq.~(\ref{temp1}) $Tr[\mathbf{J}(\vec{x})]$ is the trace ($Tr$)
of the Fisher information matrix.

Fisher information matrix, $\mathbf{J}(\vec{\phi})$, was
introduced in the twenties~\cite{pennino} and gives a different
information compared to Shannon entropy. Consider a system with a
PD that depends on a vector of parameters $\vec{\phi}$:
$p(\vec{\phi}, \vec{x})$. We can wonder which is the information
contained in a statistical sampling about the value of
$\vec{\phi}$, assuming that the real values are unknown. This
information is not related to the estimation of $\vec{\phi}$, but
more to its variance. Estimation theory states that the best
possible $\vec{\phi}$ estimator has a covariance matrix equal to
the inverse of the Fisher information matrix. This is the result
of Cramer-Rao theorem~\cite{jun}.

A particular case of Fisher information matrix is of great
relevance, when $\vec{\phi}$ reduces to a set of location
parameters~\cite{pennino}, $p(\vec{\phi}, \vec{x})$ = $p(\vec{x} -
\vec{\phi})$. In this case Fisher information matrix components
are equal to:
\begin{equation}\label{fish2}
    J_{ij}(\vec{x}) = \int \frac{1}{p(\vec{x})}
    \left (
    \frac{\partial p(\vec{x})}{\partial x_i}
    \frac{\partial p(\vec{x})}{\partial x_j}
    \right ) d\vec{x},
\end{equation}
which is the Fisher information matrix used in eq.~(\ref{temp1}). The De Bruijn's identity holds for many important PD (i.e. Gaussian PD) but not for all.
Further generalizations of this relation between entropy derivative and statistical operators is currently under investigation \cite{guo,park}.

Starting from the relation between entropy and
temperature we can link $T_{\theta}$ to the inverse
of the Fisher information matrix:
\begin{eqnarray}
  \frac{\partial H(\vec{x})}{\partial \bar{E}} &=& \frac{\partial H(\vec{x})}{\partial \epsilon}
  \frac{\partial \epsilon}{\partial \bar{E}}  = \frac{Tr[\mathbf{J}(\vec{x})]}{2}\frac{\partial \epsilon}{\partial
  \bar{E}} = \frac{1}{kT_{\theta}}.
\label{fish3}
\end{eqnarray}

In this work we further extend this definition of thermodynamic
temperature to conditional entropy:
\begin{eqnarray}
  \frac{\partial H(\vec{x}|\vec{y})}{\partial \bar{E}} &=& \frac{\partial H(\vec{x},\vec{y})}{\partial \epsilon}
  \frac{\partial \epsilon}{\partial \bar{E}} - \frac{\partial H(\vec{y})}{\partial \epsilon}
  \frac{\partial \epsilon}{\partial \bar{E}} \nonumber \\
  \frac{\partial H(\vec{x}|\vec{y})}{\partial \bar{E}} &=& \frac{Tr[\mathbf{J}(\vec{x},\vec{y})]}{2}
  \frac{\partial \epsilon}{\partial \bar{E}}, \label{fish4}
\end{eqnarray}
which leads to the thermodynamic temperature equation:
\begin{equation}\label{fish6}
    T_{\theta} = \frac{2}{k (Tr[\mathbf{J}(\vec{x},\vec{y})])}\frac{\partial \bar{E}}{\partial
    \epsilon}.
\end{equation}
This definition of thermodynamic temperature reduces to the
conventional one if the PD is the Boltzmann, but can be different
in other cases, like non-equilibrium conditions ($Neq$) and
equilibrium with information sharing ($Ieq$).

The generalization of both entropy and thermodynamic temperature
allows to extend the concept of free energy potential for the two
cases under discussion ($Ieq$ and $Neq$). It is possible to
demonstrate, after some manipulations, that in
general~\cite{sanistrava}:
\begin{equation}\label{ene1}
    \frac{\partial \bar{E}}{\partial \epsilon} = \frac{1}{2} Tr \left [ \left \langle
    \frac{\partial^{2} \mathcal{E}}{\partial x_i \partial x_j}
    \right \rangle \right ],
\end{equation}
where $\langle$ ... $\rangle$ denotes averaging respect to the PD.
This derivation remains the same also when the PD is a non
equilibrium distribution $p(\vec{x},\vec{y})$, because the
perturbation of $\vec{z}$ affects only the $\vec{x}$ random
variable vector.

To evaluate the Fisher information matrix we start making a
transformation for the generalized joint PD in a Boltzmann shape:
\begin{equation}\label{ene3}
   p(\vec{x},\vec{y}) = \frac{e^{-\beta
   \mathcal{\tilde{E}}(\vec{x},\vec{y})}}{Z(\vec{x},\vec{y})}.
\end{equation}
In eq.~\ref{ene3}, $\mathcal{\tilde{E}}$ does not represent the
real energy microstate but only a functional to reproduce the
correct PD and $Z(\vec{x},\vec{y}) = \int exp(-\beta
\mathcal{\tilde{E}}(\vec{x},\vec{y})) dxdy$ is the generalized
partition function to normalize the distribution. $\beta$ is equal
to $1/kT_0$. Evaluating the components of the Fisher information
matrix $\mathbf{J}(\vec{x},\vec{y})$ we get:
\begin{equation}\label{ene2}
    J_{ij}(\vec{x},\vec{y}) = \beta \left \langle
    \frac{\partial^{2} \mathcal{\tilde{E}}(\vec{x},\vec{y})}{\partial x_i \partial x_j} \right
    \rangle.
\end{equation}

Starting from the Helmoltz free energy potential we can finally
define the generalized potential taking into account the
conditional entropy and the conditional thermodynamic temperature:
\begin{eqnarray}\label{ene4}
    F &=& \bar{E} - kT_{\theta}H(\vec{x}|\vec{y}) = \bar{E} - kT_{0} \gamma \left (H(\vec{x}|\vec{y}) \right ),
\end{eqnarray}
where the coefficient $\gamma$ is equal to:
\begin{equation}\label{ene4a}
   \gamma = \frac{ Tr \left [ \left \langle
    \frac{\partial^{2} \mathcal{E}}{\partial x_{i}^{2}} \right
    \rangle \right ]}{ Tr \left [ \left \langle
    \frac{\partial^{2} \mathcal{\tilde{E}}(\vec{x},\vec{y})}{\partial x_{i}^{2}} \right
    \rangle \right ]} = \frac{T_{\theta}}{T_0}.
\end{equation}

The free energy differences, $\Delta F_{Ieq}$ and $\Delta
F_{Neq}$, respect to the equilibrium case, $F_{eq}$, are:
\begin{eqnarray}
  \Delta F_{Ieq}  & = &  kT_0 \left ( (1 - \gamma) H(\vec{x}) + \gamma I(\vec{x} \wedge \vec{y})
  \right ), \\
  \Delta F_{Neq} & = &  \Delta \bar{E}_{Neq} +  \Delta F_{Ieq}.
\label{ene5}
\end{eqnarray}
This is the main result of the present work. Depending if we are
simply sharing information in equilibrium conditions or fully
going out of equilibrium, the variation in free energy changes. In
the first case the increase of free energy is related to
temperature variation and entropy reduction. In the second case
($Neq$), the free energy change is related to changes in the
average energy $\Delta \bar{E}_{Neq}$ also.

If the thermodynamic temperature can be approximated to the
reservoir temperature $T_0$ ($\gamma$ $\simeq$ 1) the variation in
free energy reduces to the mutual information $kT_0 I(\vec{x}
\wedge \vec{y})$~\cite{sagawa,alessio} in the equilibrium case or
$\Delta \bar{E}_{NEQ} + kT_0 I(\vec{x} \wedge \vec{y})$ in
non-equilibrium conditions. The limit $\gamma \simeq 1$ occurs when $\vec{x}$ and $\vec{y}$
are close to be independent.

In order to show an application of the formalism for $Ieq$ we
apply it to a very simple model: a single free particle moving in
one dimension. The Hamiltonian of the system $\mathcal{E}$ reduces
to $Ax^{2}$, where $A$ = $1/2m$ (with $m$ mass of the particle)
and $x$ = $p$, momentum of the particle. The equilibrium
distribution is a normal distribution with 0 mean value and a
variance $\sigma^{2}_{x}$ = $kT_0/2A$.

In the model we assume that the reservoir can probe the
momentum of the particle with a sensor. We assume also that the measurement made by the sensor is affected by noise,
than it is described by a second random variable $y$. To simplify the model also
$y$ follows a normal distribution with 0 mean value and variance $\sigma^{2}_{y}$.

Because the outcome of $y$ is correlated with the PD of $x$ the two variables follows a joint
bivariate normal distribution:
\begin{equation}\label{joint1}
    p(x,y) = \frac{ e^{-\frac{1}{2(1-\rho^{2})}
    \left ( \frac{x^{2}}{\sigma^{2}_{x}} -
    \frac{2\rho xy}{\sigma_{x}\sigma_{y}} +
    \frac{y^{2}}{\sigma^{2}_{y}} \right ) }}{2\pi\sigma_{x}\sigma_{y}
    \sqrt{1-\rho^{2}}}.
\end{equation}
where the correlation coefficient $\rho$ = $\sigma_{xy} /
\sigma_{x}\sigma_{y}$ measures the level of correlation between
the signal of the probe and the particle real momentum. $\sigma_{xy}$ is the cross-variance and the
value of $\rho^{2}$ runs between 0 (independent variables) and 1
(completely correlated variables).

The entropy of $x$, $H(x)$, is the gaussian Shannon entropy, $H(x)
= 0.5 \ln \left ( 2\pi e \sigma^{2}_{x} \right )$, with mutual
information: $I(x \wedge y) = -0.5 \ln \left ( 1 - \rho^{2} \right
)$~\cite{cover}.

Other quantities can be easily calculated, such as the average
energy, $\bar{E} = A \sigma^{2}_{x}$ and average energy
derivative $\partial \bar{E} / \partial \epsilon$ = $A$.
$\mathcal{\tilde{E}}$ can be evaluated from $p(x,y)$:
\begin{eqnarray}
    \mathcal{\tilde{E}} &=& -kT_0 \ln ( p(x,y) ) - kT_0 \ln (Z(x,y)).
\label{joint6}
\end{eqnarray}
The average of the second derivative of $\mathcal{\tilde{E}}$,
eq.~(\ref{ene2}), is then equal to:
\begin{equation}\label{joint7}
   \left \langle
    \frac{\partial^{2} \mathcal{\tilde{E}}}{\partial x^{2}} \right
    \rangle = \frac{kT_0}{(1 - \rho^2)\sigma^{2}_{x}}.
\end{equation}

\begin{figure}[htb]
\begin{center}
\includegraphics*[width=7cm, angle=0]{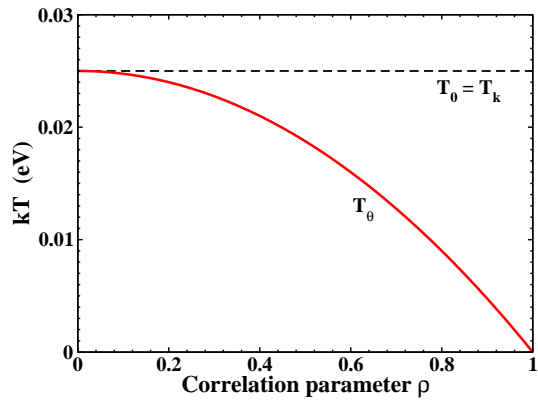}
\caption{\it{\small{(color online) Kinetic ($kT_k$ = $kT_0$) and
thermodynamic ($kT_{\theta}$) thermal energy (in eV) for different
values of the correlation coefficient $\rho$. The reservoir
temperature is assumed to be 300 K, for $kT_0$ = 0.025 eV.}}}
\label{fig:temp}
\end{center}
\end{figure}

This allows to calculate both Fisher information $J(x)$ = $\sigma_x^{-2}$ and
$J(x,y)$:
\begin{eqnarray}
    J(x,y) & = & \beta \left \langle \frac{\partial^{2} \mathcal{\tilde{E}}}{\partial x^{2}} \right
    \rangle = \frac{1}{(1 - \rho^2)\sigma^{2}_{x}},
\label{joint8}
\end{eqnarray}
which leads to $\gamma$ equals to:
\begin{eqnarray}
    \gamma & = &  \frac{2A(1 - \rho^2)\sigma^{2}_{x}}{kT_0} = (1 - \rho^2) = \frac{T_{\theta}}{T_0}.
\label{joint9}
\end{eqnarray}
The thermodynamic temperature is reduced respect to $T_0$ by increasing the
correlation (see figure~\ref{fig:temp}). The average energy on the
contrary is not affected by the correlation and then kinetic temperature $T_k$ = $T_0$. Because the heat
and work exchanged depend on different temperatures (eq.
\ref{tempeff}), in this particular case we obtain that  $\Delta Q$
= -$\Delta W$ $\propto$ ($T_{0} - T_{\theta}$). This means, as
should be, that in equilibrium condition, even when information is
acquired, the total energy flux
$\Delta Q$ + $\Delta W$ is zero. However, we observe that the
single fluxes are not zero, in particular the system absorbs heat
($\Delta Q$ $>$ 0) and makes mechanical work to the reservoir ($\Delta W$ $<$
0).

The explanation of the example is the following. Using the information about the momentum of the particle,
the reservoir can extract mechanical work from it. For example using a small mill, that can be rotated by the particle. In equilibrium
conditions the impacts between particle and mill do not produce net mechanical work,
because there are equal impacts in one direction and in the opposite. However, if the reservoir can estimate the dynamical state
of the particle with more accuracy, then can use the mill to extract work when the particle moves in one direction and block the mill when the particle
comes from the opposite direction. The impact transfers energy from the particle to the reservoir in form of a net mechanical work.
The particle cools down because its average energy is reduced by the series of impacts. However, the particle is also in thermal contact with the reservoir,
if the temperature of the particle gets lower than the reservoir, a heat transfer occurs increasing again the average energy of the particle.

So, even if the net flux of energy is zero, the
entropy flux is not. The system absorbs energy with high entropy
(heat) and gives back to the reservoir energy with a lower entropy
(work). This is consistent with the fact that, thanks to the
information due to correlation with the microstate, the
system behaves with respect to the reservoir more and more like a
deterministic system. Its internal average energy can be exploited
to make work. In fact, in the limit of perfect correlation
$T_{\theta}$ = 0 and $F$ = $\bar{E}$.

It is important to stress that this is not implying any violation
of the second principle of thermodynamics. The reduction of
entropy performed in the system occurs only thanks to the
information obtained. The harvesting and
elaboration of this information requires work made at the expense
of the reservoir. The variation of entropy due to this elaboration
more than compensate the local reduction in entropy in the system.
Then the total entropy variation, system plus reservoir, is still
non negative.

In this work a generalization of the free energy potential has
been presented. This formalism makes use of informational
theory concepts in order to extend the thermodynamic theory
to systems which share information with the reservoir or are in
steady state non-equilibrium. The core point is the concept of
mutual information and conditional Fisher information, which allow
to generalize both entropy and thermodynamic temperature
definitions, respectively. A simple model, describing a single
free particle, has been discussed showing the effect of shared
information over the thermodynamics between system and reservoir.
Increasing the correlation between measure and microstate allows to consider
the system more and more as deterministic. Then its
internal energy can be entirely converted into work.



\begin{thebibliography}{99}

\bibitem{crooksthesis} G. E. Crooks, Excursions in Statistical Dynamics, PhD Thesis,
University of California at Berkley, 1999.

\bibitem{jarzynski4} C. Jarzynski, Non-equilibrium equality for free energy differences, {\em Phys. Rev. Lett.}, vol. 78, 2690 (1997).

\bibitem{crooks4} D. A. Sivak, G. E. Crooks, Near-equilibrium measurements of nonequilibrium free
energy, {\em Phys. Rev. Lett.}, vol. 108, 150601 (2012).

\bibitem{boksen} E. Boksenbojm, B. Wynants and C. Jarzynski, Nonequilibrium thermodynamics
at the microscale: work relations and the second law, {\em
arXiv:1002.1230v1}, (2010).

\bibitem{muschik2} W. Muschik, Aspects of non-equilibrium thermodynamics, Singapore (World Scientific), (1990).

\bibitem{crooks2} G. E. Crooks, Entropy production fluctuation theorem and the non-equilibrium work relation for
free energy differences, {\em Phys. Rev. E}, vol. 60, 2721 (1999).

\bibitem{crooks3} G. E. Crooks, Path-ensemble averages in system driven far from equilibrium, {\em Phys. Rev. E}, vol. 61, 2361 (2000).

\bibitem{zubarev2} D. N. Zubarev et al., Statistical Mechanics of Nonequilibrium Processes, (Berlin: Akademie), (1996).

\bibitem{luzzi1} R. Luzzi et al, Statistical Foundations of Irreversible
Thermodynamics, (Leipzig: Teubner), (2001).




\bibitem{diventra} Y. Dubi, M. Di Ventra, Colloquium: Heat flow and thermoelectricity in atomic and molecular
junctions, {\em Rev. Mod. Phys.}, vol. 83, 131 (2011).

\bibitem{tempneq} J. C. Vasquez and D. Jou, Temperature in non-equilibrium states: a review
of open problems and current proposals, {\em Rep. Prog. Phys.},
vol. 66, 1937 (2004).

\bibitem{sagawa2} T. Sagawa and M. Ueda, {\em Phys. Rev. E}, 85, 021104 (2012).
\bibitem{sagawa3} T. Sagawa and M. Ueda, {\em Phys. Rev. Lett.}, 109, 180602 (2012).


\bibitem{alessio} A. Gagliardi and A. Di Carlo,{\em Phys. A},
vol. 391, 6337 (2012).


\bibitem{huang} K. Huang, "Statistical Mechanics (2nd ed.)", Wyley
(1987).

\bibitem{sanistrava} K. R. Narayanan and A. R. Sanistrava,{\em Phys. Rev. E},
vol. 85, 031151 (2012).

\bibitem{cover} T. M. Cover and J. A. Thomas, "Elements of information theory", Wyley (2006).






\bibitem{jaynes} E. T. Jaynes, Information theory and statistical mechanics, {\em Phys. Rev. A},
vol. 106, 620 (1957).

\bibitem{jaynes2} E. T. Jaynes, Information theory and statistical mechanics - II, {\em Phys. Rev.
A}, vol. 108, 171 (1957).

\bibitem{brill} L. Brillouin, Science and Information Theory, Mineola,
N.Y.: Dover, 2004.

\bibitem{merhav} N. Merhav, Statistical Physics and Information theory, {\em
Foundation and Trends in Communication and Information Theory},
vol. 6, 1-212 (2009).

\bibitem{bagci} G. B. Bagci, The physical meaning of Renyi relative entropies,
{\em arXiv:cond-mat/0703008v1}, (2007).




\bibitem{landauer} R. Landauer, Irreversibility and Heat Generation
in the Computing Process, {\em IBM J. Res. Develop.}, Vol. 5, No.
3, 183(1961).

\bibitem{jarzynski3} J. Horowitz and C. Jarzynski, An illustrative example of the relationship between dissipation
and relative entropy, {\em arXiv:0901.0576v1}, (2009).


\bibitem{kullback} S. Kullback, Information theory and statistics, Dover (1978).

\bibitem{pennino} F. Pennini, A.R. Plastino, A. Plastino, {\em Phys. A}, vol. 258, 446 (1998).

\bibitem{jun} J. Shao, "Mathematical Statistics", New York:
Springer (1998).


\bibitem{guo} D. Guo, S. Shamai and S. Verdu, {\em IEEE Trans. Information Theory}, vol. 51, 1261 (2005).

\bibitem{park} S. Park, E. Serpedin and K. Qaraqe, arXiv:1202.0015v4 (2012).



\bibitem{sagawa} T. Sagawa and M. Ueda, {\em Phys. Rev. Lett.}, 104, 090602 (2010).

\bibitem{footnote} The PD of the components of $\vec{z}$ are assumed to be
independent and identically normal distributed with 0 mean value
and unitary variance.


\end{thebibliography}

\end{document}